\PassOptionsToPackage{unicode}{hyperref}
\PassOptionsToPackage{hyphens}{url}
\PassOptionsToPackage{dvipsnames,svgnames,x11names}{xcolor}
\documentclass[
  12pt]{article}
\usepackage{xcolor}
\usepackage{amsmath,amssymb}
\usepackage{iftex}
\ifPDFTeX
  \usepackage[T1]{fontenc}
  \usepackage[utf8]{inputenc}
  \usepackage{textcomp} % provide euro and other symbols
\else % if luatex or xetex
  \usepackage{unicode-math} % this also loads fontspec
  \defaultfontfeatures{Scale=MatchLowercase}
  \defaultfontfeatures[\rmfamily]{Ligatures=TeX,Scale=1}
\fi
\usepackage{lmodern}
\ifPDFTeX\else
\fi
\IfFileExists{upquote.sty}{\usepackage{upquote}}{}
\IfFileExists{microtype.sty}{% use microtype if available
  \usepackage[]{microtype}
  \UseMicrotypeSet[protrusion]{basicmath} % disable protrusion for tt fonts
}{}
\makeatletter
\@ifundefined{KOMAClassName}{% if non-KOMA class
  \IfFileExists{parskip.sty}{%
    \usepackage{parskip}
  }{% else
    \setlength{\parindent}{0pt}
    \setlength{\parskip}{6pt plus 2pt minus 1pt}}
}{% if KOMA class
  \KOMAoptions{parskip=half}}
\makeatother
\makeatletter
\ifx\paragraph\undefined\else
  \let\oldparagraph\paragraph
  \renewcommand{\paragraph}{
    \@ifstar
      \xxxParagraphStar
      \xxxParagraphNoStar
  }
  \newcommand{\xxxParagraphStar}[1]{\oldparagraph*{#1}\mbox{}}
  \newcommand{\xxxParagraphNoStar}[1]{\oldparagraph{#1}\mbox{}}
\fi
\ifx\subparagraph\undefined\else
  \let\oldsubparagraph\subparagraph
  \renewcommand{\subparagraph}{
    \@ifstar
      \xxxSubParagraphStar
      \xxxSubParagraphNoStar
  }
  \newcommand{\xxxSubParagraphStar}[1]{\oldsubparagraph*{#1}\mbox{}}
  \newcommand{\xxxSubParagraphNoStar}[1]{\oldsubparagraph{#1}\mbox{}}
\fi
\makeatother

\usepackage{color}
\usepackage{fancyvrb}

\DefineVerbatimEnvironment{Highlighting}{Verbatim}{commandchars=\\\{\}}
\usepackage{framed}
\definecolor{shadecolor}{RGB}{241,243,245}
\newenvironment{Shaded}{\begin{snugshade}}{\end{snugshade}}

\newcommand{\AttributeTok}[1]{\textcolor[rgb]{0.40,0.45,0.13}{#1}}

\newcommand{\ConstantTok}[1]{\textcolor[rgb]{0.56,0.35,0.01}{#1}}

\newcommand{\DecValTok}[1]{\textcolor[rgb]{0.68,0.00,0.00}{#1}}

\newcommand{\FloatTok}[1]{\textcolor[rgb]{0.68,0.00,0.00}{#1}}
\newcommand{\FunctionTok}[1]{\textcolor[rgb]{0.28,0.35,0.67}{#1}}

\newcommand{\NormalTok}[1]{\textcolor[rgb]{0.00,0.23,0.31}{#1}}

\newcommand{\OtherTok}[1]{\textcolor[rgb]{0.00,0.23,0.31}{#1}}

\newcommand{\StringTok}[1]{\textcolor[rgb]{0.13,0.47,0.30}{#1}}

\usepackage{longtable,booktabs,array}
\usepackage{calc} % for calculating minipage widths
\usepackage{etoolbox}
\makeatletter
\patchcmd\longtable{\par}{\if@noskipsec\mbox{}\fi\par}{}{}
\makeatother
\IfFileExists{footnotehyper.sty}{\usepackage{footnotehyper}}{\usepackage{footnote}}
\makesavenoteenv{longtable}
\usepackage{graphicx}
\makeatletter
\newsavebox\pandoc@box
\newcommand*\pandocbounded[1]{% scales image to fit in text height/width
  \sbox\pandoc@box{#1}%
  \Gscale@div\@tempa{\textheight}{\dimexpr\ht\pandoc@box+\dp\pandoc@box\relax}%
  \Gscale@div\@tempb{\linewidth}{\wd\pandoc@box}%
  \ifdim\@tempb\p@<\@tempa\p@\let\@tempa\@tempb\fi% select the smaller of both
  \ifdim\@tempa\p@<\p@\scalebox{\@tempa}{\usebox\pandoc@box}%
  \else\usebox{\pandoc@box}%
  \fi%
}
\def\fps@figure{htbp}
\makeatother

\providecommand{\tightlist}{%
  \setlength{\itemsep}{0pt}\setlength{\parskip}{0pt}}

\usepackage[]{natbib}
\usepackage{makecell}
\usepackage{tabularx}
\usepackage{float}
\usepackage{etoolbox}

\date{}

\AtBeginEnvironment{figure}{\spacingset{1}}
\AtBeginEnvironment{table}{\spacingset{1}\renewcommand{\arraystretch}{1.15}}
\AtBeginEnvironment{longtable}{\spacingset{1}\renewcommand{\arraystretch}{1.15}}
\AtBeginEnvironment{Highlighting}{\spacingset{1.25}}
\AtBeginEnvironment{verbatim}{\spacingset{1}}

\makeatletter
\let\asaoriginalmaketitle\maketitle
\renewcommand{\maketitle}{%
  \let\asaoriginalaffiliationreset\asaaffiliationreset
  \renewcommand{\asaaffiliationreset}{%
    \normalsize\itshape\asaoriginalaffiliationreset%
  }%
  \patchcmd{\@author}
    {and \vspace{0.2em}\\}
    {\rule{0pt}{1em}\normalsize and\\[0.5em]}
    {}{}%
  \patchcmd{\@author}
    {and \vspace{0.2em}\\}
    {\rule{0pt}{1em}\normalsize and\\[0.5em]}
    {}{}%
  \asaoriginalmaketitle
  \let\asaaffiliationreset\asaoriginalaffiliationreset
}
\makeatother
\makeatletter
\@ifpackageloaded{caption}{}{\usepackage{caption}}
\AtBeginDocument{%
\ifdefined\contentsname
  \renewcommand*\contentsname{Table of contents}
\else
  \newcommand\contentsname{Table of contents}
\fi
\ifdefined\listfigurename
  \renewcommand*\listfigurename{List of Figures}
\else
  \newcommand\listfigurename{List of Figures}
\fi
\ifdefined\listtablename
  \renewcommand*\listtablename{List of Tables}
\else
  \newcommand\listtablename{List of Tables}
\fi
\ifdefined\figurename
  \renewcommand*\figurename{Figure}
\else
  \newcommand\figurename{Figure}
\fi
\ifdefined\tablename
  \renewcommand*\tablename{Table}
\else
  \newcommand\tablename{Table}
\fi
}
\@ifpackageloaded{float}{}{\usepackage{float}}
\floatstyle{ruled}
\@ifundefined{c@chapter}{\newfloat{codelisting}{h}{lop}}{\newfloat{codelisting}{h}{lop}[chapter]}
\floatname{codelisting}{Listing}

\makeatother
\makeatletter
\@ifpackageloaded{caption}{}{\usepackage{caption}}
\@ifpackageloaded{subcaption}{}{\usepackage{subcaption}}
\makeatother
\usepackage{bookmark}
\IfFileExists{xurl.sty}{\usepackage{xurl}}{} % add URL line breaks if available
\hypersetup{
  pdftitle={conformalForecast: An R package for sequential conformal prediction in time series},
  pdfauthor={Xiaoqian Wang; Daniele Girolimetto},
  pdfkeywords={Conformal prediction, Prediction
intervals, Calibration, Time series forecasting, Open-source software},
  colorlinks=true,
  linkcolor={blue},
  filecolor={Maroon},
  citecolor={Blue},
  urlcolor={Blue},
  pdfcreator={LaTeX via pandoc}}

\begin{document}
\def\spacingset#1{\renewcommand{\baselinestretch}%
{#1}\small\normalsize} \spacingset{1}

\renewcommand*{\arraystretch}{0.5} % Specify row height in a table globally

% Build affiliation lines from whichever normalized Quarto fields are present.
\newif\ifasaaffiliationhasitem
\newcommand{\asaaffiliationreset}{\asaaffiliationhasitemfalse}
\newcommand{\asaaffiliationitem}[1]{%
  \ifasaaffiliationhasitem, \fi#1\asaaffiliationhasitemtrue%
}

% Record whether the metadata identifies at least one corresponding author.
\newif\ifasacorrespondingauthor
\asacorrespondingauthorfalse

%%%%%%%%%%%%%%%%%%%%%%%%%%%%%%%%%%%%%%%%%%%%%%%%%%%%%%%%%%%%%%%%%%%%%%%%%%%%%%

\title{\bf conformalForecast: An R package for sequential conformal
prediction in time
series\ifasacorrespondingauthor\footnotemark[2]\else\footnotemark[1]\fi}
\author{\parbox{0.96\textwidth}{\centering
{\large
Xiaoqian Wang\textsuperscript{a}, Daniele
Girolimetto\textsuperscript{b}}\\[0.55em]
{\normalsize\itshape
 \textsuperscript{a}Center for Forecasting Science, Academy of
Mathematics and Systems Science, Chinese Academy of
Sciences, Beijing 100190, China\\[0.1em] \textsuperscript{b}Department
of Statistical Sciences, University of Padova, Padova 35121, Italy}%
}}
\maketitle
\begingroup
\ifasacorrespondingauthor
\renewcommand{\thefootnote}{\fnsymbol{footnote}}
\footnotetext[1]{Corresponding author\par
\textit{Email addresses:} \nolinkurl{xiaoqian.wang@amss.ac.cn} (Xiaoqian
Wang), \nolinkurl{daniele.girolimetto@unipd.it} (Daniele Girolimetto)}
\else
\renewcommand{\thefootnote}{}
\footnotetext{\textit{Email addresses:} \nolinkurl{xiaoqian.wang@amss.ac.cn} (Xiaoqian
Wang), \nolinkurl{daniele.girolimetto@unipd.it} (Daniele Girolimetto)}
\fi
\endgroup
\begingroup
\renewcommand{\thefootnote}{\fnsymbol{footnote}}
\ifasacorrespondingauthor
\footnotetext[2]{We thank Rob J. Hyndman for his valuable guidance and
constructive suggestions during the development of the conformalForecast
package, as well as for his helpful comments on an earlier version of
this paper. Xiaoqian Wang is supported by the National Natural Science
Foundation of China (No.~72601399) and the Presidential Foundation of
the Academy of Mathematics and Systems Science, Chinese Academy of
Sciences, China (No.~E555930101).}
\else
\footnotetext[1]{We thank Rob J. Hyndman for his valuable guidance and
constructive suggestions during the development of the conformalForecast
package, as well as for his helpful comments on an earlier version of
this paper. Xiaoqian Wang is supported by the National Natural Science
Foundation of China (No.~72601399) and the Presidential Foundation of
the Academy of Mathematics and Systems Science, Chinese Academy of
Sciences, China (No.~E555930101).}
\fi
\endgroup

\begin{abstract}
Reliable uncertainty quantification is central to modern forecasting.
Conformal prediction offers model-agnostic intervals with coverage
guarantee under exchangeability, but existing software often targets
general prediction problems or individual time series procedures rather
than integrated workflows for multistep forecasting applications. We
introduce \textbf{conformalForecast}, an open-source R package for
constructing, updating, and evaluating conformal prediction intervals
for multistep time series forecasts. The package separates point
forecasting from calibration, accepts forecasts from diverse sources
through a common interface, and preserves horizon-specific forecast
errors and calibration states. It implements several conformal
calibration schemes spanning empirical-quantile, feedback-based, and
dependence-aware approaches, together with tools for assessing coverage,
interval width, and interval scores over time and across horizons.
Fitted objects can be extended as new observations arrive while
retaining previously issued intervals and avoiding recomputation at
earlier forecast origins. An application to short-term electricity price
forecasting across five Australian regions illustrates the workflow.
Aggregate coverage is generally close to the nominal level, while
rolling coverage and interval efficiency vary across methods and
horizons. In the reported experiment, incremental updating reproduces
full-reconstruction intervals under identical forecasting and
calibration settings while substantially reducing computation.
\textbf{conformalForecast} provides a reproducible and extensible
framework linking conformal methodology with sequential forecasting
practice.
\end{abstract}

\noindent%
{\it Keywords:} Conformal prediction, Prediction
intervals, Calibration, Time series forecasting, Open-source software
\vfill

\newpage
\spacingset{1.25}

\section{Introduction}\label{sec-intro}

Forecasting requires not only accurate point predictions but also
reliable quantification of their uncertainty. In short-term wholesale
electricity markets, for example, a nominal 90\% price prediction
interval supports hedging and operational planning only if it is well
calibrated, with empirical coverage close to 90\%, and sharp enough to
distinguish different levels of price risk. Conformal prediction
\citep{papadopoulos2002, vovk2005} provides a general framework for
constructing such intervals. Under exchangeability, it offers
finite-sample, distribution-free marginal coverage guarantees
\citep{shafer2008, lei2018}, and it can be applied on top of any point
forecast, whether from a statistical/machine learning model, or a
judgmental process, without specifying a predictive distribution.

Time series pose a fundamental challenge for this framework. Serial
dependence, nonstationarity, and distributional shifts violate
exchangeability, so the classical guarantee need not hold. One line of
research extends conformal prediction beyond exchangeability under
additional conditions on the form of the shift or dependence
\citep{tibshirani2019, chernozhukov2018, xu2021, barber2023}. A
complementary line adapts calibration online as observations arrive,
without modeling how the distribution evolves. Adaptive conformal
inference adjusts the miscoverage level in response to past coverage
errors \citep{gibbs2021}, and later developments include expert
aggregation \citep{zaffran2022}, adaptive regret minimization
\citep{bhatnagar2023}, dynamically tuned step sizes \citep{gibbs2024},
PID-based threshold updates \citep{angelopoulos2023}, and multistep
forecasting with dependence across horizons \citep{wang2024}. These
methods turn calibration into a sequential task, and this sequential
structure makes their implementation a forecasting-specific software
problem.

Forecasting practice imposes requirements that extend beyond a generic
regression interface. Forecasts are issued repeatedly as the forecast
origin advances, often at several horizons, and their errors become
observable only as outcomes are realized. A conformal forecasting
workflow should therefore (i) accept point forecasts from any source,
(ii) preserve the joint origin--horizon structure of the forecast
errors, (iii) update calibration as new observations arrive without
repeating earlier computations, and (iv) evaluate interval performance
over time and across horizons.

Existing open-source software meets these requirements only in part.
General-purpose tools such as \textbf{pintervals}
\citep{randahl2026pintervals}, \textbf{marginaleffects}
\citep{arel-bundock2024}, and \textbf{probably} \citep{kuhn2025probably}
within \textbf{tidymodels} workflows \citep{kuhn2020tidymodels} in R;
\textbf{crepes} \citep{bostrom2024}, \textbf{puncc} \citep{mendil2023},
and \textbf{TorchCP} \citep{huang2025} in Python; and
\textbf{ConformalPrediction.jl} \citep{altmeyer2022} in Julia, target
regression and classification problems. Other R packages address
specific model classes or response structures, including
\textbf{conformalbayes} \citep{mccartan2025} for Bayesian models,
\textbf{conformalInference.fd} \citep{diquigiovanni2022fd} and
\textbf{conformalInference.multi} \citep{diquigiovanni2025multi} for
functional and multivariate responses, and \textbf{LCP} \citep{guan2023}
for localized conformal prediction. A small number of packages address
the time series framework. In R, \textbf{modeltime}
\citep{dancho2026modeltime} adds Gaussian-residual and split-conformal
intervals to \textbf{tidymodels} workflows, whereas
\textbf{caretForecast} \citep{akay2026} constructs horizon-specific
intervals by rolling-origin calibration. In Python, \textbf{sktime}
\citep{sktime2026} wraps arbitrary forecasters with empirical and
conformal intervals. \textbf{StatsForecast}
\citep{garza2022statsforecast} and \textbf{MLForecast}
\citep{morales2026mlforecast} calibrate multistep intervals for
statistical and machine-learning forecasters, respectively, using
rolling-origin residuals. \textbf{MAPIE} \citep{cordier2023} implements
EnbPI and adaptive conformal inference through a dedicated time series
regressor, while \textbf{puncc} and \textbf{crepes} also provide
sequential procedures. Few of these combine the four requirements within
a single model-agnostic workflow, so applying and comparing conformal
methods across forecasting models and horizons still requires custom
code.

This paper presents \textbf{conformalForecast}, an open-source R package
whose design follows from these four requirements. Point forecasts may
come from many sources, including operational systems outside R, so
calibration is separated from forecasting and accepts any function that
returns a \texttt{forecast} object. Forecasts issued at successive
origins are stored with their errors in a common origin--horizon
representation that respects the horizon-dependent delay with which
errors become observable; five calibration schemes, including the
multistep approach of \citet{wang2024}, operate on it through a single
interface. As new observations arrive, fitted objects are extended
rather than rebuilt: with the forecasting function and calibration
settings unchanged, updating reproduces the intervals of a full
reconstruction and leaves previously issued intervals untouched.
Finally, since calibration quality varies over time and with the
horizon, coverage, interval width, and interval scores are reported
overall, over rolling windows, and by horizon. The resulting objects
integrate with the \textbf{forecast} ecosystem
\citep{Hyndman2008-forecast, hyndman2026}.

We illustrate these design choices in an operational setting: the
calibration of short-term electricity price forecasts published by the
Australian Energy Market Operator for the five regions of the National
Electricity Market. The point forecasts are produced outside R and come
without prediction intervals; they are issued repeatedly at horizons
from 15 to 60 minutes, and their intervals are updated as new prices
become available. The package is available from CRAN, with the
development version at
\url{https://github.com/xqnwang/conformalForecast}; the analyses in this
paper use version 0.2.0.

The remainder of the paper is organized as follows.
Section~\ref{sec-method} reviews the online conformal prediction methods
implemented in the package. Section~\ref{sec-software} presents the
design and implementation of \textbf{conformalForecast}.
Section~\ref{sec-application} demonstrates the workflow in short-term
electricity price forecasting. Section~\ref{sec-conclusion} concludes.

\section{Time series conformal prediction}\label{sec-method}

\subsection{Sequential multistep forecasting
framework}\label{sec-framework}

\begin{figure}[h]

\centering{

\includegraphics[width=1\linewidth,height=\textheight,keepaspectratio]{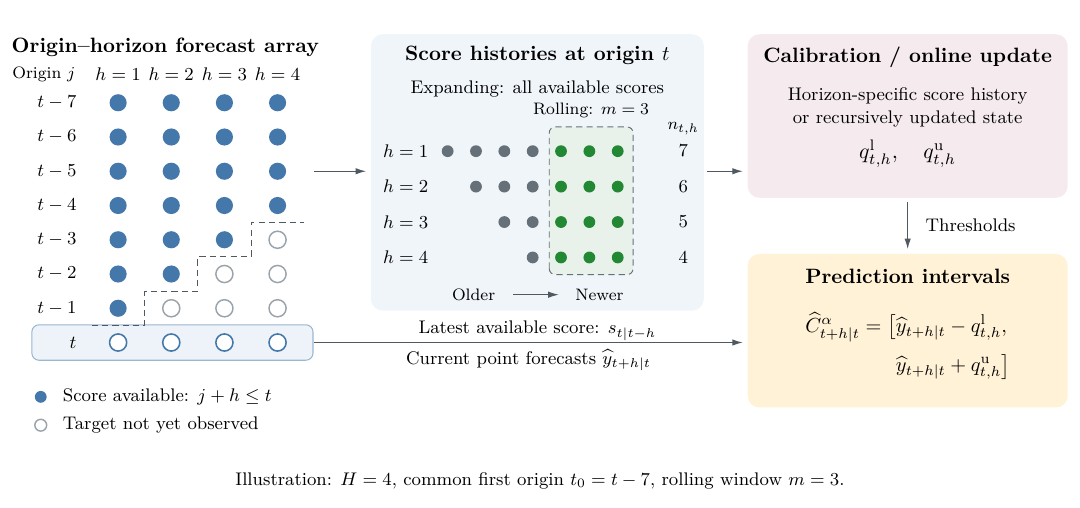}

}

\caption{\label{fig-framework}Sequential multistep forecasting and
conformal calibration.}

\end{figure}

Let \(\{y_t\}_{t=1}^T\) denote a univariate time series. At forecast
origin \(t\), a forecasting method produces point forecasts
\(\widehat{y}_{t+h\mid t}\) for horizons \(h=1,\ldots,H\), with error
\(e_{t+h\mid t}=y_{t+h}-\widehat{y}_{t+h\mid t}\) once the target is
observed. Repeating this procedure over successive origins
\citep{Hyndman2021-fpp3} gives the origin--horizon array in
Figure~\ref{fig-framework}. The conformal layer aligns forecasts with
their targets and does not require the internal specification of the
forecasting model.

At origin \(t\), an \(h\)-step forecast issued at origin \(j\) can
contribute a calibration score only when \(j+h\leq t\). With initial
origin \(t_0\), the expanding history is \[
\mathcal{S}^{\mathrm{exp}}_{t,h}
=
\left(s_{j+h\mid j}\right)_{j=t_0}^{\,t-h},
\] and is empty when \(t-h<t_0\). Feedback arrives with an \(h\)-period
delay, so longer horizons have shorter histories under a common starting
origin. A rolling history retains the most recent \(m\) available
scores, whereas an expanding history retains all of them. Neither should
be confused with the data window used to estimate the point forecasting
model.

Symmetric calibration uses absolute error
\(s_{t+h\mid t}=|e_{t+h\mid t}|\). Asymmetric calibration instead uses
\(s^{\mathrm{l}}_{t+h\mid t}=-e_{t+h\mid t}\) and
\(s^{\mathrm{u}}_{t+h\mid t}=e_{t+h\mid t}\). For target miscoverage
\(\alpha\in(0,1)\), the relevant histories or recursively updated states
produce thresholds \(q^{\mathrm{l}}_{t,h}\) and \(q^{\mathrm{u}}_{t,h}\)
and the interval \[
\widehat{C}^{\alpha}_{t+h\mid t}
=
\left[
\widehat{y}_{t+h\mid t}-q^{\mathrm{l}}_{t,h},
\widehat{y}_{t+h\mid t}+q^{\mathrm{u}}_{t,h}
\right].
\] Symmetric calibration sets
\(q^{\mathrm{l}}_{t,h}=q^{\mathrm{u}}_{t,h}=q_{t,h}\); asymmetric
calibration allocates the total miscoverage across the two tails.

The methods below differ in how these horizon-specific score streams
determine the thresholds, but share this input--output structure with
the \textbf{conformalForecast} package in Section~\ref{sec-software}.

\subsection{Conformal calibration methods}\label{sec-cpmethods}

Coverage statements require a distinction between marginal validity and
long-run calibration. Classical conformal prediction gives finite-sample
marginal validity,
\(\Pr\{y_{t+h}\in\widehat C^\alpha_{t+h\mid t}\}\geq1-\alpha\), when the
calibration and test scores are exchangeable. For online methods, let
\(\operatorname{err}_{u\mid u-h}=\mathbf{1}\{y_u\notin\widehat C^\alpha_{u\mid u-h}\}\)
and let \(u_h\) be the first evaluated target time. Long-run validity
requires the time-average miscoverage to have limit superior at most
\(\alpha\), whereas exact long-run calibration requires \[
\lim_{T\to\infty}
\frac{1}{T-u_h+1}\sum_{u=u_h}^{T}\operatorname{err}_{u\mid u-h}
=\alpha.
\] Neither property implies conditional coverage at each origin,
coverage in every local window, or simultaneous coverage of the forecast
path.

Table \ref{tab:calibration-methods} summarizes the five calibration
schemes implemented in \textbf{conformalForecast}, which fall into three
families: empirical quantile calibration, feedback-based online
calibration, and dependence-aware multistep calibration. They differ in
the quantity that is adapted as new scores arrive and in the coverage
statement that is available, together with the conditions under which it
holds.

\begin{table}[!ht]
\centering
\caption{Calibration schemes implemented in \textbf{conformalForecast}:
quantity adapted over time, available coverage statement, and conditions
under which it holds.}
\label{tab:calibration-methods}
\begingroup
\footnotesize
\newcolumntype{Y}{>{\raggedright\arraybackslash}X}
\begin{tabularx}{\textwidth}{@{}l
  >{\hsize=1.03\hsize\linewidth=\hsize}Y
  >{\hsize=0.92\hsize\linewidth=\hsize}Y
  >{\hsize=1.05\hsize\linewidth=\hsize}Y@{}}
\toprule
Method & Adapted quantity & Coverage statement & Conditions \\
\midrule
\multicolumn{4}{@{}l}{\textit{Empirical quantile}} \\
\quad SCP & Nothing: empirical score quantile
  & Finite-sample marginal coverage
  & Exchangeable calibration and test scores \\
\quad WCP & Nothing: weighted score quantile
  & Finite-sample marginal coverage under exchangeability; otherwise a bounded coverage shortfall
  & Fixed, data-independent weights; small weighted swap total variation term for near-nominal coverage \\
\addlinespace[3pt]
\multicolumn{4}{@{}l}{\textit{Feedback-based}} \\
\quad ACP & Miscoverage level $\alpha_{t,h}$, updated from past violations
  & Long-run calibration for $h=1$; no established long-run guarantee for multistep extension
  & Fixed positive step size; boundary quantile convention \\
\quad PID & Threshold $q_{t,h}$, through tracking, integration, and scorecasting
  & Long-run calibration at each fixed forecast horizon
  & Bounded scores and baseline thresholds; required integrator saturation \\
\addlinespace[3pt]
\multicolumn{4}{@{}l}{\textit{Dependence-aware}} \\
\quad AcMCP & Threshold $q_{t,h}$, with an MA$(h-1)$ forecast-error adjustment
  & Long-run calibration at each fixed forecast horizon
  & Bounded scores and baseline thresholds; required integrator saturation \\
\bottomrule
\end{tabularx}
\endgroup
\end{table}

\subsubsection{Empirical quantile
calibration}\label{empirical-quantile-calibration}

Split conformal prediction (SCP) calibrates each horizon with an
empirical score quantile \citep{papadopoulos2002, vovk2005, lei2018}.
For the origins \(\mathcal J_{t,h}\) retained at \((t,h)\), the
corrected construction is \[
q_{t,h}
=Q_{1-\alpha_*}\!\left(
\sum_{j\in\mathcal J_{t,h}}\widetilde w_{j,t,h}\,
\delta_{s_{j+h\mid j}}
+\widetilde w_{\infty,t,h}\,\delta_{+\infty}
\right),
\] where \(Q_\tau\) is the generalized inverse quantile, \(\delta_a\) is
a unit point mass, and \(\alpha_*=\alpha\) for absolute scores or
\(\alpha/2\) for each signed tail. SCP assigns mass
\(1/(|\mathcal J_{t,h}|+1)\) to every calibration score and to
\(+\infty\), which supplies the conformal rank correction. If the
forecaster is fitted on separate training data and the calibration and
test scores are exchangeable, this construction has finite-sample
marginal validity without requiring a correctly specified forecasting
model. Calibrating each signed tail at \(\alpha/2\) gives interval
coverage of at least \(1-\alpha\) by the union bound. Temporal
dependence, distributional change, and repeated refitting can violate
score exchangeability in rolling-origin applications. Interpolated
quantiles and effective-sample-size adjustments likewise do not
automatically inherit the exact guarantee.

Weighted split conformal prediction (WCP) uses the same construction
with prespecified, data-independent weights \citep{barber2023}. Each
calibration score receives \(w_j\in[0,1]\), the point mass at
\(+\infty\) receives weight one, and the weights are normalized. The
corrected quantile remains marginally valid when the calibration and
test scores are exchangeable. More generally, under the swap-based
framework of \citet{barber2023}, the coverage shortfall is bounded by a
weighted sum of total-variation distances between the joint sequence and
its calibration--test swaps. Recency weighting can reduce the influence
of less representative scores, but does not itself guarantee validity
under dependence or distributional change. In
\textbf{conformalForecast}, WCP is obtained from \texttt{scp()} with a
user-supplied weight function. The stated theoretical results require
the corresponding exchangeability or swap conditions, admissible fixed
weights, and the corrected generalized-inverse quantile; other
weighting, quantile, and effective-sample-size choices are
implementation options described in Section~\ref{sec-software-cp}.

\subsubsection{Feedback-based online
calibration}\label{feedback-based-online-calibration}

Adaptive conformal prediction (ACP) adjusts the miscoverage level using
observed score violations \citep{gibbs2021}. For horizon \(h\), let
\(v_{t,h}=\mathbf{1}\{s_{t\mid t-h}>q_{t-h,h}\}\) indicate whether the
newly observed score exceeds its issued threshold. Then \[
\alpha_{t,h}
=\alpha_{t-1,h}+\gamma\{\alpha_*-v_{t,h}\},
\] where \(\gamma>0\) controls responsiveness, and \(\alpha_*=\alpha\)
for symmetric calibration or \(\alpha/2\) for each tail. A violation
lowers \(\alpha_{t,h}\) and typically raises the next threshold. The
original one-step ACI recursion achieves distribution-free long-run
calibration when levels outside \([0,1]\) correspond to full or empty
prediction sets; applying that one-step recursion separately to the two
tails gives interval coverage of at least \(1-\alpha\). Following
\citet{wang2024}, \textbf{conformalForecast} instead applies the update
to each delayed horizon-specific stream. This operational multistep
extension does not automatically inherit the one-step result, and
excursions outside \([0,1]\) can produce unbounded or empty intervals.

Conformal PID control instead updates thresholds directly on the score
scale \citep{angelopoulos2023}. For each horizon \(h\), the proportional
tracking state and final threshold are \[
\begin{aligned}
q^P_{t,h}
&=q^P_{t-1,h}+\eta_{t,h}(v_{t,h}-\alpha_*),\\
q_{t,h}
&=q^P_{t,h}
+r_{t,h}\!\left(\sum_{u=b_h}^{t}(v_{u,h}-\alpha_*)\right)
+\widehat s_{t+h\mid t},
\end{aligned}
\] Here \(\eta_{t,h}\) is the tracking step size, \(r_{t,h}\) is a
saturating error integrator, \(b_h\) starts the controller feedback, and
\(\widehat s_{t+h\mid t}\) is an optional predictable score forecast.
These terms provide reactive tracking, cumulative correction, and
prospective adjustment, respectively. \textbf{conformalForecast}
maintains horizon-specific states; with asymmetric scores, the
signed-error forecast enters the two tail thresholds with opposite
signs. Under the score, predictable-threshold, and integrator-saturation
conditions of \citet{angelopoulos2023}, extended to delayed multistep
feedback by \citet{wang2024}, the violation frequency at each horizon
converges to \(\alpha_*\). Separate tail calibration therefore gives
long-run interval coverage of at least \(1-\alpha\). The finite-sample
bound contains a delay term that increases with \(h\) for a fixed
evaluation length.

\subsubsection{Dependence-aware multistep
calibration}\label{dependence-aware-multistep-calibration}

Autocorrelated multistep conformal prediction (AcMCP) replaces the
generic PID scorecaster with a structured forecast-error adjustment
\citep{wang2024}. A first-order approximation for a general
non-stationary autoregressive process gives optimal \(h\)-step errors an
approximate MA\((h-1)\) representation. AcMCP therefore combines an
MA\((h-1)\) model for the available \(h\)-step errors with a regression
linking errors from the same forecast origin; unavailable within-path
errors are supplied recursively. \textbf{conformalForecast} applies this
adjustment to signed scores with separate tail thresholds. Under the
same predictable-threshold and integrator conditions as the multistep
PID construction, AcMCP achieves exact long-run calibration at each
horizon. The adjustment targets interval efficiency rather than a
stronger coverage criterion, while longer horizons require more history
and incur a larger finite-sample delay term.
Section~\ref{sec-application} compares these methods using common
rolling-origin forecasts.

\section{The R package conformalForecast}\label{sec-software}

\subsection{Package architecture}\label{sec-software-design}

The \textbf{conformalForecast} package provides an R \citep{RCoreTeam}
implementation of the methods described in Section~\ref{sec-method}. It
is available on CRAN, with source code hosted on GitHub, and is
distributed under the GPL-3 license. The package uses the S3 class
system and a data structure compatible with \textbf{forecast}
\citep{Hyndman2008-forecast}. Its core object classes inherit from
\texttt{forecast}; \textbf{conformalForecast} defines class-specific
methods for \texttt{print()}, \texttt{summary()}, and
\texttt{accuracy()}, while stored forward forecasts can be visualized
through the inherited \texttt{autoplot()} method. Additional
dependencies include \textbf{ggdist} \citep{Kay2024-ggdist} for weighted
sample quantiles, \textbf{zoo} \citep{Zeileis2005-zoo} for rolling
summaries, and \textbf{rlang} \citep{Henry2026-rlang} together with
\textbf{stats} for supporting programming and statistical utilities.

Point forecasting and conformal calibration form separate stages. At
successive forecast origins, \texttt{cvforecast()} calls a user-supplied
forecasting function that returns a \texttt{forecast} object. The
function may fit a model in R or wrap forecasts generated externally, as
illustrated in Section~\ref{sec-application}. The conformal stage
operates on the resulting origin--horizon arrays of point forecasts and
realized errors rather than on the internal representation of the
forecasting model. Consequently, the same rolling-origin forecasts and
error history can be supplied to each conformal method, holding the
point forecast input fixed when methods are compared.

\begin{figure}[h]

\centering{

\includegraphics[width=1\linewidth,height=\textheight,keepaspectratio]{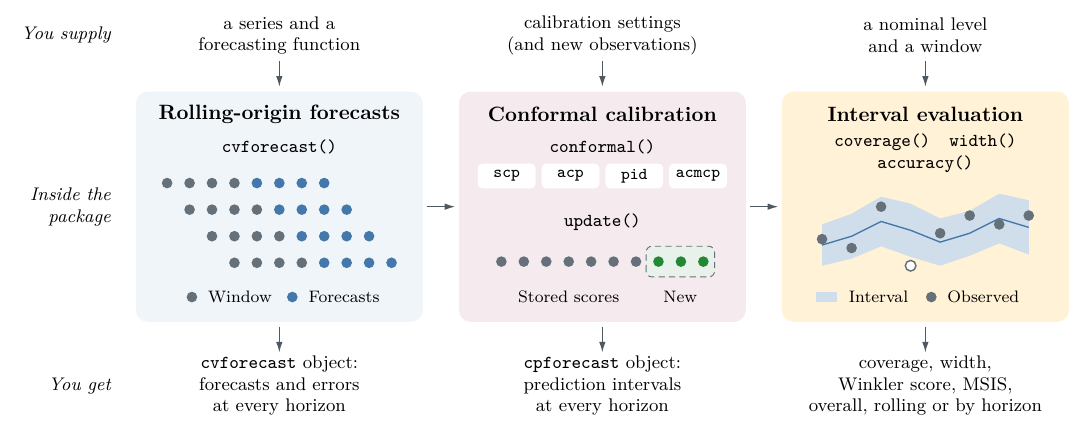}

}

\caption{\label{fig-workflow}Overview of the \textbf{conformalForecast}
workflow. \texttt{cvforecast()} applies the user-supplied forecasting
function at successive origins and records the resulting point forecasts
and errors. \texttt{conformal()} uses these errors to construct
prediction intervals, while \texttt{update()} extends an existing
calibration when new observations become available. The resulting
objects can be evaluated using \texttt{coverage()}, \texttt{width()},
and \texttt{accuracy()}.}

\end{figure}

Two core S3 classes encode these stages. A \texttt{cvforecast} object
inherits from \texttt{forecast} and stores the rolling-origin forecasts
and realized errors by horizon, together with any intervals supplied by
the forecasting function if available. A \texttt{cpforecast} object is
also a subclass of \texttt{cvforecast}; it contains the calibrated
intervals and the method-specific information needed for sequential
updating, and carries an additional class identifying the selected
method. As shown in Figure~\ref{fig-workflow}, \texttt{conformal()}
dispatches on a \texttt{cvforecast} object and returns the corresponding
method-specific \texttt{cpforecast} object. Objects from either stage
can be evaluated using \texttt{coverage()}, \texttt{width()}, and
\texttt{accuracy()}, whereas \texttt{update()} operates on a
\texttt{cpforecast} object and returns an updated object that extends
the calibration sequence when new observations become available.

\subsection{Rolling-origin forecasting}\label{sec-software-cv}

The conformal methods require rolling-origin forecast errors, which are
computed by \texttt{cvforecast()}:

\begin{Shaded}
\begin{Highlighting}[]
\FunctionTok{cvforecast}\NormalTok{(y, forecastfun, }\AttributeTok{h =} \DecValTok{1}\NormalTok{, }\AttributeTok{level =} \FunctionTok{c}\NormalTok{(}\DecValTok{80}\NormalTok{, }\DecValTok{95}\NormalTok{), }\AttributeTok{forward =} \ConstantTok{TRUE}\NormalTok{,}
           \AttributeTok{xreg =} \ConstantTok{NULL}\NormalTok{, }\AttributeTok{initial =} \DecValTok{1}\NormalTok{, }\AttributeTok{window =} \ConstantTok{NULL}\NormalTok{, ...)}
\end{Highlighting}
\end{Shaded}

It moves a forecast origin along \texttt{y} and, at each origin, applies
\texttt{forecastfun} and stores the point forecasts and the errors for
all horizons up to \texttt{h}. The forecasting function takes a
univariate time series as its first argument and accepts \texttt{h} and
\texttt{level}, together with \texttt{xreg} and \texttt{newxreg} when
exogenous predictors are used. It returns a \texttt{forecast} object
whose \texttt{mean}, \texttt{lower}, and \texttt{upper} components are
aligned with the requested horizons and coverage levels. When the
forecasting method does not provide prediction intervals, \texttt{lower}
and \texttt{upper} can instead contain missing values, as in
Section~\ref{sec-application}. Beyond this interface, the package places
no restriction on the forecasting method.

The location of the forecast origins is determined by \texttt{initial}
and \texttt{window}. When \texttt{window\ =\ NULL}, the first origin is
\texttt{initial}, and the estimation sample expands as new observations
become available. When a rolling window is specified, each origin uses
the most recent \texttt{window} observations, and the first origin is
\texttt{max(initial,\ window)}. If \texttt{forward\ =\ TRUE},
\texttt{cvforecast()} also generates forecasts from the final
observation in the series, storing these out-of-sample forecasts and any
associated intervals or fitted-model information in the \texttt{mean},
\texttt{lower}, \texttt{upper}, and \texttt{model} components. The
\texttt{level} argument supplies one or more nominal coverage levels,
which are stored in the resulting object and used in the subsequent
conformal calibration.

The result is a \texttt{cvforecast} object. Its \texttt{MEAN} and
\texttt{ERROR} components are multivariate time series indexed by target
time, with the \(h\)th column containing point forecasts and errors,
respectively, for horizon \(h\). The \texttt{LOWER} and \texttt{UPPER}
components preserve the interval bounds returned by
\texttt{forecastfun}; they are lists indexed by nominal coverage level,
each containing multivariate time series in the same layout. In the
\texttt{cpforecast} object described in Section~\ref{sec-software-cp},
the corresponding components instead contain the conformally calibrated
intervals. The forecast errors
\(e_{t+h\mid t}=y_{t+h}-\widehat y_{t+h\mid t}\) are transformed into
absolute or signed nonconformity scores during calibration, as defined
in Section~\ref{sec-framework}. The point forecasts in \texttt{MEAN}
locate the resulting intervals.

\subsection{Conformal calibration}\label{sec-software-cp}

The \texttt{conformal()} S3 generic provides a common interface to the
conformal methods in Section~\ref{sec-cpmethods}:

\begin{Shaded}
\begin{Highlighting}[]
\FunctionTok{conformal}\NormalTok{(object, }\AttributeTok{method =} \FunctionTok{c}\NormalTok{(}\StringTok{"scp"}\NormalTok{, }\StringTok{"acp"}\NormalTok{, }\StringTok{"pid"}\NormalTok{, }\StringTok{"acmcp"}\NormalTok{), ...)}
\end{Highlighting}
\end{Shaded}

For a \texttt{cvforecast} object, each value of \texttt{method} selects
the corresponding implementation function and passes \texttt{object} and
\texttt{...} unchanged, so
\texttt{conformal(fc,\ method\ =\ "scp",\ ...)} and
\texttt{scp(fc,\ ...)} are equivalent. The interface contains four such
functions: \texttt{scp()}, \texttt{acp()}, \texttt{pid()}, and
\texttt{acmcp()}. WCP is obtained by supplying a non-uniform weighting
rule through \texttt{weightfun} to \texttt{scp()}, whereas SCP uses
equal weights. Table~\ref{tbl-methods} summarizes the four
implementations and their additional arguments.

\begin{table}[H]

\caption{\label{tbl-methods}Implementations selected by the
\texttt{method} argument in \textbf{conformalForecast} and their
principal additional arguments. Arguments common to all implementations
are described in the text.}

\centering{

\begin{tabular}{llp{7.9cm}}
\toprule
\textbf{\texttt{method}} & \textbf{Function} & \textbf{Principal additional arguments}\\
\midrule
\texttt{"scp"} & \texttt{scp()} & \texttt{weightfun}, \texttt{kess}\\
\texttt{"acp"} & \texttt{acp()} & \texttt{gamma}\\
\texttt{"pid"} & \texttt{pid()} & \texttt{integrate}, \texttt{scorecast}, \texttt{scorecastfun}, \texttt{lr}, \texttt{Tg}, \texttt{delta}, \texttt{Csat}, \texttt{KI}\\
\texttt{"acmcp"} & \texttt{acmcp()} & \texttt{integrate}, \texttt{scorecast}, \texttt{lr}, \texttt{Tg}, \texttt{delta}, \texttt{Csat}, \texttt{KI}, \texttt{ma\_method}\\
\bottomrule
\end{tabular}

}

\end{table}%

The target miscoverage, \texttt{alpha}, is taken by default from the
levels stored in the object. In \texttt{scp()} and \texttt{acp()},
\texttt{ncal} specifies the initial size of an expanding calibration
history or the length of a rolling history of past nonconformity scores.
In \texttt{pid()} and \texttt{acmcp()}, it defines the initial
calibration and scorecasting burn-in period. When
\texttt{rolling\ =\ TRUE}, it also determines the length of the trailing
histories used for scorecasting and learning-rate calculations.
Otherwise, these histories expand as observations become available.
Because \texttt{acmcp()} constructs scorecasts recursively across
horizons, longer horizons require additional history before all
scorecaster inputs are available. The \texttt{symmetric} argument
selects absolute error scores or separate signed scores in
\texttt{scp()}, \texttt{acp()}, and \texttt{pid()}, whereas
\texttt{acmcp()} supports only the latter. For \texttt{scp()} and
\texttt{acp()}, \texttt{quantiletype} selects one of the nine estimators
implemented by \texttt{ggdist::weighted\_quantile()}. The default, type
1, matches the generalized-inverse definition in
Section~\ref{sec-cpmethods}. The finite-sample result stated there for
SCP and WCP relies on this corrected quantile construction together with
the corresponding assumptions and weighting conditions. Other quantile
types are implementation options and do not automatically inherit that
result.

Function-specific arguments control weighting, online adaptation, and
nonconformity-score forecasting. When the calibration history contains
\(m\) nonconformity scores, \texttt{scp()} calls \texttt{weightfun} with
\(m+1\) and applies the returned vector of fixed, non-data-dependent
weights to the historical scores and the additional \(+\infty\) value.
This construction allows recent historical scores to receive greater
weight than older ones while retaining an explicit weight for the added
point mass. When weighted scores are used, setting
\texttt{kess\ =\ TRUE} bases the quantile calculation on Kish's
effective sample size rather than on the number of available values. As
noted in Section~\ref{sec-cpmethods}, this adjustment is an
implementation option and does not automatically inherit the exact
weighted conformal guarantee. The step size used by \texttt{acp()} to
update the miscoverage level over time is determined by \texttt{gamma}.

For \texttt{pid()}, \texttt{lr} supplies the initial scale for the
proportional quantile-tracking update. The learning rate used in the
recursion is subsequently scaled by the range of the available signed or
absolute errors. The arguments \texttt{integrate}, \texttt{KI},
\texttt{Csat}, \texttt{Tg}, and \texttt{delta} configure the integral
component. The derivative component is specified through
\texttt{scorecast} and \texttt{scorecastfun}. The latter accepts a
univariate time series and a forecast horizon and returns a
\texttt{forecast} object whose \texttt{mean} component supplies the
nonconformity-score forecasts. Most of these controller arguments are
also available in \texttt{acmcp()}, whose structured error-based
scorecaster is defined internally and therefore does not require
\texttt{scorecastfun}. The additional argument \texttt{ma\_method}
determines how the moving-average component is estimated.

All four implementation functions and calls through the
\texttt{conformal()} S3 generic return a \texttt{cpforecast} object with
the same \texttt{MEAN} and \texttt{ERROR} components as its
\texttt{cvforecast} input, while \texttt{LOWER} and \texttt{UPPER}
contain the calibrated intervals in the same by-horizon layout. By
preserving a common structure across conformal methods, the package
provides a uniform interface for inspecting, comparing, and processing
calibrated forecasts.

\subsection{Evaluating prediction intervals}\label{sec-software-eval}

The intervals are summarized by three functions, each of which accepts
either a \texttt{cvforecast} or a \texttt{cpforecast} object, so that a
conformal method can be compared with the intervals of the underlying
model on the same terms. The \texttt{accuracy()} call below shows
simplified usage and omits the optional \texttt{x}, \texttt{CV}, and
\texttt{period} arguments.

\begin{Shaded}
\begin{Highlighting}[]
\FunctionTok{coverage}\NormalTok{(object, ..., }\AttributeTok{level =} \DecValTok{95}\NormalTok{, }\AttributeTok{window =} \ConstantTok{NULL}\NormalTok{, }\AttributeTok{na.rm =} \ConstantTok{FALSE}\NormalTok{)}
\FunctionTok{width}\NormalTok{(object, ..., }\AttributeTok{level =} \DecValTok{95}\NormalTok{, }\AttributeTok{includemedian =} \ConstantTok{FALSE}\NormalTok{, }\AttributeTok{window =} \ConstantTok{NULL}\NormalTok{,}
      \AttributeTok{na.rm =} \ConstantTok{FALSE}\NormalTok{)}
\FunctionTok{accuracy}\NormalTok{(object, }\AttributeTok{measures =}\NormalTok{ interval\_measures, }\AttributeTok{byhorizon =} \ConstantTok{FALSE}\NormalTok{, ...)}
\end{Highlighting}
\end{Shaded}

The \texttt{coverage()} function returns empirical coverage for each
horizon as both a mean over the evaluation sample and a sequence of
observation-level indicators. The \texttt{width()} function returns the
interval widths together with their mean for each horizon and, when
\texttt{includemedian\ =\ TRUE}, their median. Supplying \texttt{window}
adds rolling mean coverage and rolling mean width, together with rolling
median width when requested. These rolling summaries are used in
Section~\ref{sec-application} to examine changes in local coverage and
interval width.

Accuracy measures come from \texttt{accuracy()}, which computes them
from a list supplied by the user. The package exports two such lists:
\texttt{point\_measures}, with ME, MAE, MSE, RMSE, MPE, MAPE, MASE and
RMSSE, and \texttt{interval\_measures}, with the Winkler score
\citep{Winkler1972-interval} and the mean scaled interval score
\citep{Makridakis2020-m4}. Users may instead provide a list of functions
that accept the relevant inputs supplied by \texttt{accuracy()}. With
\texttt{byhorizon\ =\ TRUE}, each measure is reported separately by
horizon, allowing changes in interval performance across forecast
horizons to be examined.

\subsection{Updating with new observations}\label{sec-software-update}

Rebuilding the pipeline every time an observation arrives gives the
right answer at an unnecessary cost: the model is refitted at origins
that were already fitted, and intervals that have already been reported
are computed again. Both are avoided by \texttt{update()}:

\begin{Shaded}
\begin{Highlighting}[]
\FunctionTok{update}\NormalTok{(object, new\_data, forecastfun, }\AttributeTok{new\_xreg =} \ConstantTok{NULL}\NormalTok{, ...)}
\end{Highlighting}
\end{Shaded}

It appends \texttt{new\_data}, computes forecasts only for the newly
available origins, and extends the calibration using the stored score
histories or recursive state. Because the forecasting function is not
stored in the object, each update call requires \texttt{forecastfun} to
be supplied again. It should be the same function used in the original
call to \texttt{cvforecast()}.

During an update, the conformal method and its calibration settings are
recovered from the fitted object and kept unchanged. Modifying
parameters such as \texttt{ncal}, \texttt{lr}, or \texttt{KI} therefore
requires a new calibration rather than an update. When exogenous
predictors were used in the original fit, \texttt{new\_xreg} is required
and contains one row per new observation. These rows extend the stored
predictor matrix over the periods immediately following it. When
\texttt{forward\ =\ TRUE}, that matrix can already contain future
predictor values, so \texttt{new\_xreg} continues from the end of the
stored matrix rather than necessarily aligning row by row with
\texttt{new\_data}. With the same forecasting function and calibration
settings, the updated object produces the same intervals as a complete
reconstruction of the pipeline on the extended series, while leaving
previously issued intervals unchanged. As a result, the computational
cost is driven mainly by the number of new observations rather than by
the length of the full series.

\section{Application}\label{sec-application}

\subsection{Data source and preprocessing}\label{sec-app-data}

We illustrate the package on wholesale electricity prices from the
Australian National Electricity Market (NEM), operated by the Australian
Energy Market Operator (AEMO) \citep{AEMO2026-mmsdm}. AEMO's P5MIN
process produces twelve five-minute regional price forecasts for the
next hour, with a new run every five minutes. At each selected
quarter-hour reference time, we aggregate forecasts from a single run
into four non-overlapping 15-minute forecasts and apply the same
aggregation to realized prices. The reference times identify the
forecast schedule rather than the exact publication times. The complete
data cover April 2026, with 2,880 selected reference times per region.
The main analysis uses 1--29 April, while 30 April is reserved for the
updating experiment in Section~\ref{sec-app-update}.

This application demonstrates how the package can calibrate forecasts
produced by an external system. The point forecasts are published by
AEMO rather than estimated within our analysis. The P5MIN data used here
contain no associated prediction intervals or predictive distribution.
The application therefore illustrates the model-independent interface
described in Section~\ref{sec-software-design}.

\begin{figure}

\centering{

\includegraphics[width=1\linewidth,height=\textheight,keepaspectratio]{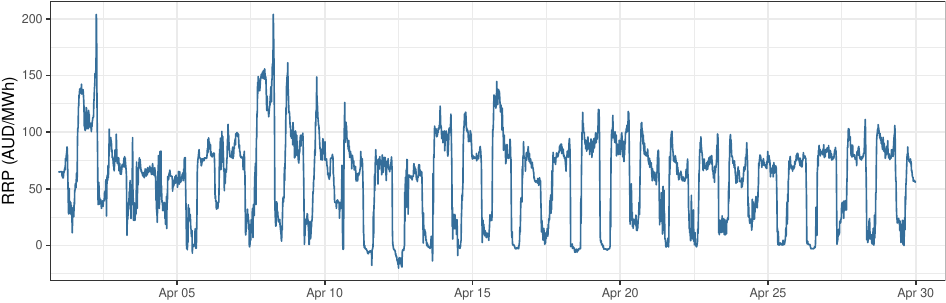}

}

\caption{\label{fig-nsw-prices}Realized 15-minute electricity prices in
New South Wales over the analysis sample, 1 to 29 April 2026.}

\end{figure}

Figure~\ref{fig-nsw-prices} shows the New South Wales series for 1--29
April 2026. Prices range from \(-20\) to \(204\) AUD/MWh, with a median
of \(67.5\). Negative prices are genuine market outcomes rather than
data errors. Price variability changes considerably over time, with
relatively stable periods interrupted by sharp intraday movements. The
corresponding forecast errors exhibit time-varying dispersion and
horizon-dependent asymmetry. These features motivate interval methods
that can adapt their width over time and calibrate the two tails
separately.

\subsection{Calibration setup and implementation}\label{sec-app-setup}

We first define a forecasting function that retrieves the four
aggregated forecasts associated with each rolling origin. We then use
\texttt{cvforecast()} to compute the corresponding forecast errors and
apply the five conformal methods, implemented through four calibration
functions, to the nonconformity scores derived from these errors.

In this application, \texttt{forecastfun} does not estimate a model. It
identifies the current origin from the time index, retrieves the four
15-minute forecasts constructed from the corresponding AEMO P5MIN run,
and returns them as the \texttt{mean} component of a \texttt{forecast}
object. The \texttt{lower} and \texttt{upper} components are filled with
missing values because the P5MIN data used here include no intervals and
the conformal calibration uses forecast errors, rather than intervals
from the underlying forecasting system. The matrix of aggregated
forecasts and the start of the series are handed to \texttt{forecastfun}
as extra arguments, which \texttt{cvforecast()} passes on through its
\texttt{...}.

The call to \texttt{cvforecast()} walks the forecast origin along the
series and records the error of each aggregated forecast at each
horizon:

\begin{Shaded}
\begin{Highlighting}[]
\NormalTok{fc }\OtherTok{\textless{}{-}} \FunctionTok{cvforecast}\NormalTok{(y, }\AttributeTok{forecastfun =}\NormalTok{ forecastfun, }\AttributeTok{h =} \DecValTok{4}\NormalTok{, }\AttributeTok{level =} \DecValTok{90}\NormalTok{,}
                 \AttributeTok{window =} \DecValTok{192}\NormalTok{, }\AttributeTok{forecasts =}\NormalTok{ forecasts,}
                 \AttributeTok{series\_start =}\NormalTok{ series\_start)}
\end{Highlighting}
\end{Shaded}

We analyze the four constructed horizons at a nominal level of 90\%. The
rolling window of 192 observations is two days of quarter-hourly data.
The analysis series runs from 1 to 29 April and contains 2,784
observations. The final day is held back for the update in
Section~\ref{sec-app-update}. After the initial window,
\texttt{cvforecast()} produces forecasts at 2,593 origins, including the
final out-of-sample origin. The first is at 23:45 on 2 April. Each
conformal method uses 192 realized forecast errors per horizon before
issuing its first intervals on 5 April, leaving 2,400, 2,398, 2,396 and
2,394 evaluated intervals at horizons 1--4, respectively.

The object \texttt{fc} now carries the point forecasts and the errors,
and calibrating intervals from them is a single call. All five methods
use signed errors and calibrate the lower and upper tails separately.
For AcMCP:

\begin{Shaded}
\begin{Highlighting}[]
\NormalTok{acmcpfc }\OtherTok{\textless{}{-}} \FunctionTok{conformal}\NormalTok{(fc, }\AttributeTok{method =} \StringTok{"acmcp"}\NormalTok{, }\AttributeTok{ncal =} \DecValTok{192}\NormalTok{, }\AttributeTok{rolling =} \ConstantTok{TRUE}\NormalTok{,}
                     \AttributeTok{lr =} \FloatTok{0.005}\NormalTok{, }\AttributeTok{Tg =} \DecValTok{1000}\NormalTok{, }\AttributeTok{KI =} \DecValTok{100}\NormalTok{)}
\end{Highlighting}
\end{Shaded}

We set \texttt{ncal} equal to the rolling-window length in
\texttt{cvforecast()}, so the calibration history and forecasting input
window each contain 192 observations. The result is a
\texttt{cpforecast} object, and the three evaluation functions of
Section~\ref{sec-software-eval} take it directly:

\begin{Shaded}
\begin{Highlighting}[]
\FunctionTok{accuracy}\NormalTok{(acmcpfc, }\AttributeTok{measures =}\NormalTok{ interval\_measures, }\AttributeTok{level =} \DecValTok{90}\NormalTok{,}
         \AttributeTok{byhorizon =} \ConstantTok{TRUE}\NormalTok{)}
\FunctionTok{coverage}\NormalTok{(acmcpfc, }\AttributeTok{level =} \DecValTok{90}\NormalTok{, }\AttributeTok{window =} \DecValTok{100}\NormalTok{)}
\FunctionTok{width}\NormalTok{(acmcpfc, }\AttributeTok{level =} \DecValTok{90}\NormalTok{, }\AttributeTok{includemedian =} \ConstantTok{TRUE}\NormalTok{, }\AttributeTok{window =} \DecValTok{100}\NormalTok{)}
\end{Highlighting}
\end{Shaded}

We evaluate local interval performance using a rolling window of 100
valid intervals, corresponding to 25 hours of quarter-hourly
observations. The rolling coverage and median width, together with
summaries over the complete evaluation sample, are reported in
Section~\ref{sec-app-nsw}. The across-horizon coverage reported there is
obtained by pooling the available entries in the indicator matrix
returned by \texttt{coverage()}.

The remaining four methods are applied to the same \texttt{cvforecast}
object by changing \texttt{method} where applicable and supplying the
relevant tuning arguments:

\begin{itemize}
\tightlist
\item
  SCP and WCP are fitted with \texttt{method\ =\ "scp"}. WCP
  additionally supplies \texttt{weightfun} with exponentially decaying,
  non-data-dependent weights and decay factor \(0.99\).
\item
  ACP is fitted with \texttt{method\ =\ "acp"} and adaptation step
  \texttt{gamma\ =\ 0.005}.
\item
  Conformal PID is fitted with \texttt{method\ =\ "pid"} and uses a
  naive nonconformity-score forecast supplied through
  \texttt{scorecastfun}, together with \texttt{lr\ =\ 0.005},
  \texttt{Tg\ =\ 1000}, and \texttt{KI\ =\ 100}. The same values of
  \texttt{lr}, \texttt{Tg}, and \texttt{KI} are used for AcMCP.
\end{itemize}

These settings are held fixed across regions and horizons to keep the
comparison focused on a common software workflow.

\subsection{Results for New South Wales}\label{sec-app-nsw}

\begin{figure}

\centering{

\includegraphics[width=1\linewidth,height=\textheight,keepaspectratio]{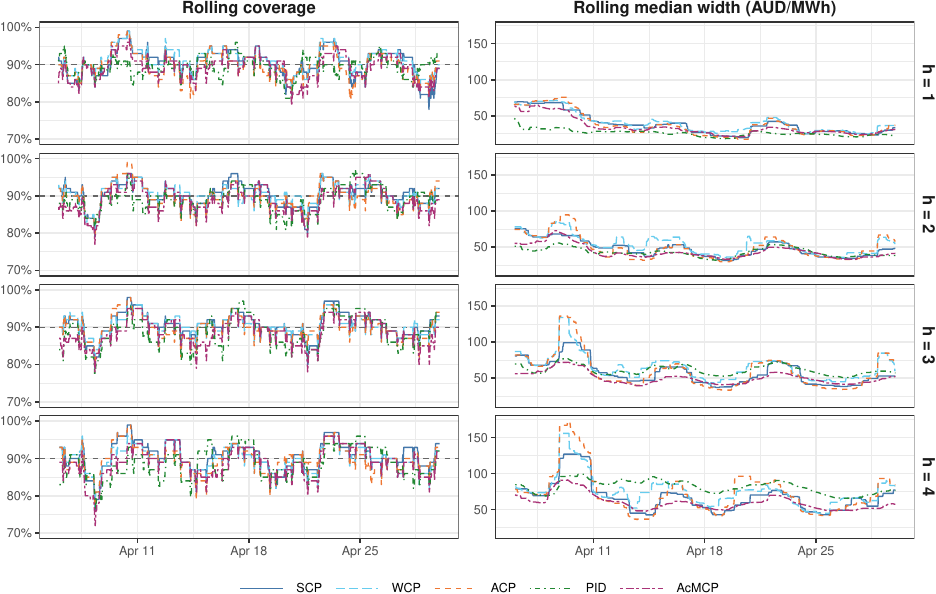}

}

\caption{\label{fig-nsw-rolling}Rolling performance of the 90\%
prediction intervals for New South Wales, shown separately by forecast
horizon. The left panels report mean coverage over the most recent 100
valid intervals, with the nominal level indicated by the dashed line,
while the right panels report rolling median interval width over the
same complete window.}

\end{figure}

Figure~\ref{fig-nsw-rolling} shows the rolling summaries returned by
\texttt{coverage()} and \texttt{width()} after 100 valid intervals have
accumulated for each method and horizon. The left panel shows local
coverage, and the right one shows sharpness. The coverage estimates over
the complete evaluation sample reported in Table~\ref{tbl-nsw} are close
to 90\%, whereas local coverage fluctuates appreciably over time. The
right panels show substantial differences in interval width over time
and across horizons, and the relative ordering of the methods varies
with horizon.

Table~\ref{tbl-nsw} reports coverage, interval width, and interval
scores both by forecast horizon and after pooling across horizons. Small
differences in empirical coverage remain. Interval width and the proper
scoring rules therefore provide additional information about the
relative performance of the methods.

PID with a naive \texttt{scorecaster} produces the narrowest mean
intervals at the shortest horizon, but its intervals widen more rapidly
than those of the other methods as the forecast horizon increases. At
the longest horizon, PID produces the widest intervals. This pattern is
consistent with the horizon-specific fitting of its scorecaster, which
does not use dependence between forecast errors across horizons.

AcMCP produces the second-narrowest mean intervals at the first two
horizons and the narrowest at the two longer horizons. This pattern is
consistent with its use of dependence across forecast horizons, as
described in Section~\ref{sec-method}. Although AcMCP also has the
narrowest pooled mean width, SCP attains the lowest pooled Winkler score
and MSIS, as well as the lowest values of both scores at the two longer
horizons. These results illustrate that narrower intervals do not
necessarily imply better overall performance when sharpness and coverage
are considered together.

\begin{table}

\caption{\label{tbl-nsw}New South Wales, 90\% prediction intervals.
Empirical coverage, mean and median interval width, Winkler score and
MSIS, by forecast horizon and pooled over horizons
(\(h = 1, \dots, 4\)). Widths and Winkler scores are in AUD/MWh.}

\centering{

\resizebox{\linewidth}{!}{\setlength{\tabcolsep}{1pt}%
\begin{tabular}{@{}l@{\hspace{4pt}}|rrrrr@{\hspace{4pt}}|rrrrr@{}}
\toprule
 & \multicolumn{10}{c}{\textbf{Conformal method}}\\
\textbf{Measure} & \makebox[3.5em][r]{\footnotesize \textbf{SCP}} & \makebox[3.5em][r]{\footnotesize \textbf{WCP}} & \makebox[3.5em][r]{\footnotesize \textbf{ACP}} & \makebox[3.5em][r]{\footnotesize \textbf{PID}} & \makebox[3.5em][r]{\footnotesize \textbf{AcMCP}} & \makebox[3.5em][r]{\footnotesize \textbf{SCP}} & \makebox[3.5em][r]{\footnotesize \textbf{WCP}} & \makebox[3.5em][r]{\footnotesize \textbf{ACP}} & \makebox[3.5em][r]{\footnotesize \textbf{PID}} & \makebox[3.5em][r]{\footnotesize \textbf{AcMCP}}\\
\midrule
 & \multicolumn{5}{c@{\hspace{4pt}}|}{\textit{$h = 1$}} & \multicolumn{5}{c}{\textit{$h = 2$}}\rule[-0.9ex]{0pt}{0pt}\\
Coverage & \makebox[3.5em][r]{0.905} & \makebox[3.5em][r]{0.911} & \makebox[3.5em][r]{0.898} & \makebox[3.5em][r]{0.895} & \makebox[3.5em][r]{0.891} & \makebox[3.5em][r]{0.905} & \makebox[3.5em][r]{0.905} & \makebox[3.5em][r]{0.900} & \makebox[3.5em][r]{0.891} & \makebox[3.5em][r]{0.889}\\
Mean width & \makebox[3.5em][r]{38.4} & \makebox[3.5em][r]{40.7} & \makebox[3.5em][r]{38.2} & \makebox[3.5em][r]{26.2} & \makebox[3.5em][r]{34.7} & \makebox[3.5em][r]{48.2} & \makebox[3.5em][r]{54.8} & \makebox[3.5em][r]{50.7} & \makebox[3.5em][r]{42.8} & \makebox[3.5em][r]{43.7}\\
Median width & \makebox[3.5em][r]{36.5} & \makebox[3.5em][r]{36.6} & \makebox[3.5em][r]{33.5} & \makebox[3.5em][r]{24.9} & \makebox[3.5em][r]{29.9} & \makebox[3.5em][r]{47.1} & \makebox[3.5em][r]{52.6} & \makebox[3.5em][r]{47.5} & \makebox[3.5em][r]{41.3} & \makebox[3.5em][r]{40.9}\\
Winkler & \makebox[3.5em][r]{58.1} & \makebox[3.5em][r]{57.8} & \makebox[3.5em][r]{58.0} & \makebox[3.5em][r]{44.8} & \makebox[3.5em][r]{56.4} & \makebox[3.5em][r]{81.0} & \makebox[3.5em][r]{87.0} & \makebox[3.5em][r]{85.5} & \makebox[3.5em][r]{76.1} & \makebox[3.5em][r]{81.8}\\
MSIS & \makebox[3.5em][r]{2.51} & \makebox[3.5em][r]{2.50} & \makebox[3.5em][r]{2.51} & \makebox[3.5em][r]{1.94} & \makebox[3.5em][r]{2.44} & \makebox[3.5em][r]{3.53} & \makebox[3.5em][r]{3.80} & \makebox[3.5em][r]{3.73} & \makebox[3.5em][r]{3.34} & \makebox[3.5em][r]{3.57}\\
\midrule
 & \multicolumn{5}{c@{\hspace{4pt}}|}{\textit{$h = 3$}} & \multicolumn{5}{c}{\textit{$h = 4$}}\rule[-0.9ex]{0pt}{0pt}\\
Coverage & \makebox[3.5em][r]{0.906} & \makebox[3.5em][r]{0.904} & \makebox[3.5em][r]{0.900} & \makebox[3.5em][r]{0.886} & \makebox[3.5em][r]{0.887} & \makebox[3.5em][r]{0.907} & \makebox[3.5em][r]{0.900} & \makebox[3.5em][r]{0.900} & \makebox[3.5em][r]{0.886} & \makebox[3.5em][r]{0.890}\\
Mean width & \makebox[3.5em][r]{57.5} & \makebox[3.5em][r]{65.6} & \makebox[3.5em][r]{60.5} & \makebox[3.5em][r]{60.9} & \makebox[3.5em][r]{50.5} & \makebox[3.5em][r]{67.8} & \makebox[3.5em][r]{76.6} & \makebox[3.5em][r]{74.4} & \makebox[3.5em][r]{81.2} & \makebox[3.5em][r]{60.1}\\
Median width & \makebox[3.5em][r]{53.1} & \makebox[3.5em][r]{62.3} & \makebox[3.5em][r]{54.4} & \makebox[3.5em][r]{59.8} & \makebox[3.5em][r]{48.4} & \makebox[3.5em][r]{67.5} & \makebox[3.5em][r]{72.8} & \makebox[3.5em][r]{67.4} & \makebox[3.5em][r]{81.0} & \makebox[3.5em][r]{57.4}\\
Winkler & \makebox[3.5em][r]{99.2} & \makebox[3.5em][r]{109.5} & \makebox[3.5em][r]{106.7} & \makebox[3.5em][r]{110.3} & \makebox[3.5em][r]{101.2} & \makebox[3.5em][r]{118.2} & \makebox[3.5em][r]{132.4} & \makebox[3.5em][r]{128.9} & \makebox[3.5em][r]{147.3} & \makebox[3.5em][r]{121.4}\\
MSIS & \makebox[3.5em][r]{4.33} & \makebox[3.5em][r]{4.79} & \makebox[3.5em][r]{4.66} & \makebox[3.5em][r]{4.86} & \makebox[3.5em][r]{4.43} & \makebox[3.5em][r]{5.15} & \makebox[3.5em][r]{5.77} & \makebox[3.5em][r]{5.61} & \makebox[3.5em][r]{6.43} & \makebox[3.5em][r]{5.28}\\
\midrule
 & \multicolumn{5}{c@{\hspace{4pt}}|}{\textit{$h = 1, \dots, 4$}} & \multicolumn{5}{c}{}\rule[-0.9ex]{0pt}{0pt}\\
Coverage & \makebox[3.5em][r]{0.905} & \makebox[3.5em][r]{0.905} & \makebox[3.5em][r]{0.899} & \makebox[3.5em][r]{0.890} & \makebox[3.5em][r]{0.889} & \makebox[3.5em][r]{} & \makebox[3.5em][r]{} & \makebox[3.5em][r]{} & \makebox[3.5em][r]{} & \makebox[3.5em][r]{}\\
Mean width & \makebox[3.5em][r]{53.0} & \makebox[3.5em][r]{59.4} & \makebox[3.5em][r]{55.9} & \makebox[3.5em][r]{52.8} & \makebox[3.5em][r]{47.2} & \makebox[3.5em][r]{} & \makebox[3.5em][r]{} & \makebox[3.5em][r]{} & \makebox[3.5em][r]{} & \makebox[3.5em][r]{}\\
Median width & \makebox[3.5em][r]{49.6} & \makebox[3.5em][r]{57.2} & \makebox[3.5em][r]{50.1} & \makebox[3.5em][r]{52.3} & \makebox[3.5em][r]{46.5} & \makebox[3.5em][r]{} & \makebox[3.5em][r]{} & \makebox[3.5em][r]{} & \makebox[3.5em][r]{} & \makebox[3.5em][r]{}\\
Winkler & \makebox[3.5em][r]{89.1} & \makebox[3.5em][r]{96.6} & \makebox[3.5em][r]{94.7} & \makebox[3.5em][r]{94.5} & \makebox[3.5em][r]{90.2} & \makebox[3.5em][r]{} & \makebox[3.5em][r]{} & \makebox[3.5em][r]{} & \makebox[3.5em][r]{} & \makebox[3.5em][r]{}\\
MSIS & \makebox[3.5em][r]{3.88} & \makebox[3.5em][r]{4.21} & \makebox[3.5em][r]{4.12} & \makebox[3.5em][r]{4.14} & \makebox[3.5em][r]{3.93} & \makebox[3.5em][r]{} & \makebox[3.5em][r]{} & \makebox[3.5em][r]{} & \makebox[3.5em][r]{} & \makebox[3.5em][r]{}\\
\bottomrule
\end{tabular}}

}

\end{table}%

\subsection{Updating and computational
performance}\label{sec-app-update}

\begin{figure}[h]

\centering{

\includegraphics[width=1\linewidth,height=\textheight,keepaspectratio]{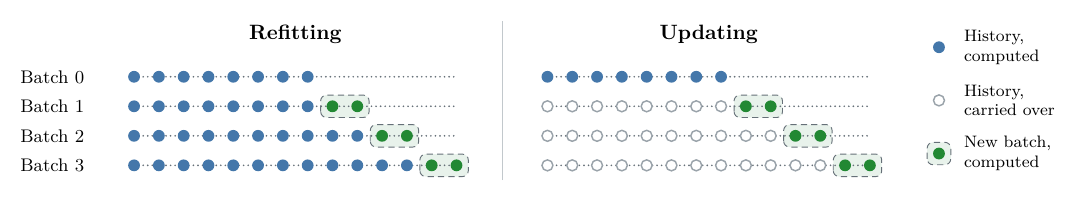}

}

\caption{\label{fig-update-scheme}The two ways of keeping a conformal
fit current as prices arrive. Each row is the series held at that
arrival, with one dot per forecast origin, and its length grows by one
batch each time. Batch 0 represents the common initial calibration cost.
From then on, refitting recomputes every origin, while \texttt{update()}
computes only the origins the new batch has added and carries the rest
over.}

\end{figure}

In this quarter-hourly application, rebuilding the pipeline whenever a
new observation becomes available repeats computations performed at
earlier forecast origins. We compare two strategies for incorporating
new observations, namely rebuilding the complete pipeline and extending
the existing objects, as shown in Figure~\ref{fig-update-scheme}.
Refitting appends the new prices to the series and runs the whole
pipeline again, so the work at each arrival grows with the accumulated
history. Extending the fit with \texttt{update()} modifies only the
origins added by the new observations, so its work is driven by the size
of the batch.

The second route is a single call. We take the AcMCP fit of
Section~\ref{sec-app-setup}, hand it the price realized at 00:00 on 30
April as \texttt{new\_price}, and supply \texttt{forecastfun} with the
arguments it needs:

\begin{Shaded}
\begin{Highlighting}[]
\NormalTok{updated\_fit }\OtherTok{\textless{}{-}} \FunctionTok{update}\NormalTok{(acmcpfc, }\AttributeTok{new\_data =}\NormalTok{ new\_price,}
                      \AttributeTok{forecastfun =}\NormalTok{ forecastfun, }\AttributeTok{forecasts =}\NormalTok{ forecasts,}
                      \AttributeTok{series\_start =}\NormalTok{ series\_start)}
\end{Highlighting}
\end{Shaded}

\begin{figure}

\centering{

\includegraphics[width=1\linewidth,height=\textheight,keepaspectratio]{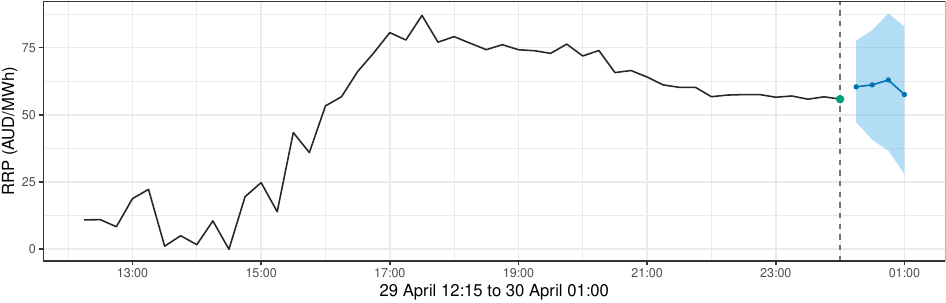}

}

\caption{\label{fig-nsw-update}AcMCP point forecasts and 90\% prediction
intervals for New South Wales after updating the fitted object with the
price observed at 00:00 on 30 April 2026. The green point identifies the
observation supplied to \texttt{update()}, the dashed line marks the new
forecast origin, and the shaded band represents the prediction intervals
for the following four quarter-hours.}

\end{figure}

The method and calibration settings are retrieved from the fitted
object, the intervals already issued for April are left as they are, and
only the new origin is calibrated. Since the fit was built with
\texttt{forward\ =\ TRUE}, the out-of-sample origin moves to the end of
the extended series, so the updated object carries point forecasts and
90\% intervals for the four quarter-hours that follow.
Figure~\ref{fig-nsw-update} plots them against the preceding twelve
hours. The intervals widen with the forecast horizon. Because AcMCP
calibrates the lower and upper tails separately using signed
nonconformity scores, the resulting intervals need not be centered on
the point forecasts, as described in Section~\ref{sec-method}.

The saving this brings can be measured on the day the analysis sample
holds back. The starting point is the calibration of
Section~\ref{sec-app-setup}, fitted on all 2,784 observations up to 29
April. It is batch 0 of Table~\ref{tbl-timing} and a cost both routes
pay. The 96 prices of 30 April then arrive in 24 hourly batches of 4,
the first of which opens with the price behind
Figure~\ref{fig-nsw-update}. The timing comparison uses one
representative fit for each implementation function, namely SCP, ACP,
PID, and AcMCP. WCP is not timed separately because it invokes the same
\texttt{scp()} update path with a different weighting rule. At each
arrival, these four fits are extended in place and, for comparison, the
corresponding pipeline is rebuilt. With the forecasting function and
calibration settings held fixed, both routes give the same intervals as
described in Section~\ref{sec-software-update}, so the table compares
computational cost at equal output.

The timings were obtained on a MacBook Pro with a 14-core Apple M4 Pro
processor (48 GB of memory) and the version of
\textbf{conformalForecast} reported in Section~\ref{sec-intro}. They are
intended as illustrative measurements for this implementation and
computing environment rather than hardware-independent benchmarks.

\begin{table}[!ht]

\caption{\label{tbl-timing}Elapsed seconds for the initial fit and for
five of the hourly arrivals that follow it, over 30 April in New South
Wales. Batch 0 is the initial fit and is the same for both routes. From
batch 1 on, refitting reruns the cross-validation and recalibrates the
SCP, ACP, PID, and AcMCP fits over the whole history, while updating
extends the fitted objects in place and never repeats the
cross-validation, which is why its CV column is empty.}

\centering{

\begin{tabular}{ccccccccccc}
\toprule
& \multicolumn{5}{c}{\textbf{Refitting}} & \multicolumn{5}{c}{\textbf{Updating}}\\
\cmidrule(lr){2-6} \cmidrule(lr){7-11}
\textbf{Batch} & \textbf{CV} & \textbf{SCP} & \textbf{ACP} & \textbf{PID} & \textbf{AcMCP} & \textbf{CV} & \textbf{SCP} & \textbf{ACP} & \textbf{PID} & \textbf{AcMCP}\\
\midrule
0 & 0.73 & 2.01 & 1.91 & 6.80 & 27.82 & 0.73 & 2.01 & 1.91 & 6.80 & 27.82\\
1 & 0.72 & 1.96 & 1.92 & 6.92 & 28.15 & -- & 0.05 & 0.01 & 0.01 & 0.05\\
6 & 0.72 & 1.94 & 1.98 & 6.98 & 28.41 & -- & 0.04 & 0.01 & 0.01 & 0.05\\
12 & 0.56 & 2.04 & 1.96 & 6.12 & 28.00 & -- & 0.05 & 0.01 & 0.01 & 0.05\\
18 & 0.68 & 2.04 & 1.95 & 5.67 & 29.40 & -- & 0.05 & 0.01 & 0.01 & 0.05\\
24 & 0.64 & 2.00 & 1.98 & 5.44 & 28.51 & -- & 0.05 & 0.01 & 0.01 & 0.05\\
\bottomrule
\end{tabular}

}

\end{table}%

Rebuilding the pipeline costs about 39 seconds per hourly batch, with
recalibration of AcMCP accounting for most of this time. The refitting
time remains approximately stable over this one-day experiment, during
which the accumulated history increases by only 3.4\%. Extending the
same four fits costs about 0.11 seconds per batch and is approximately
346 times faster in this experiment. Because refitting revisits the
complete history while updating processes only the new batch, this
advantage is expected to increase with the length of the history. Over
the day, the total computation time falls from 943 seconds to 2.7
seconds, including the removal of 15 seconds spent repeating the
rolling-origin forecasts. Both routes incur the 39 seconds required for
the initial fit only once. The full set of 24 batches is reported in the
supplementary material.

\subsection{Comparison across NEM regions}\label{sec-app-regions}

We then apply the complete pipeline to the remaining NEM regions without
changing the calibration settings. The results pooled across forecast
horizons are reported in Table~\ref{tbl-regions}, while the
corresponding horizon-specific results are provided in the supplementary
material.

\begin{table}

\caption{\label{tbl-regions}The five NEM regions, 90\% prediction
intervals pooled over horizons \(h = 1, \dots, 4\). Empirical coverage,
mean and median interval width, Winkler score and MSIS. Widths and
Winkler scores are in AUD/MWh. An infinite entry means the method issued
at least one unbounded interval.}

\centering{

\resizebox{\linewidth}{!}{\setlength{\tabcolsep}{1pt}%
\begin{tabular}{@{}l@{\hspace{4pt}}|rrrrr@{\hspace{4pt}}|rrrrr@{}}
\toprule
 & \multicolumn{10}{c}{\textbf{Conformal method}}\\
\textbf{Measure} & \makebox[3.5em][r]{\footnotesize \textbf{SCP}} & \makebox[3.5em][r]{\footnotesize \textbf{WCP}} & \makebox[3.5em][r]{\footnotesize \textbf{ACP}} & \makebox[3.5em][r]{\footnotesize \textbf{PID}} & \makebox[3.5em][r]{\footnotesize \textbf{AcMCP}} & \makebox[3.5em][r]{\footnotesize \textbf{SCP}} & \makebox[3.5em][r]{\footnotesize \textbf{WCP}} & \makebox[3.5em][r]{\footnotesize \textbf{ACP}} & \makebox[3.5em][r]{\footnotesize \textbf{PID}} & \makebox[3.5em][r]{\footnotesize \textbf{AcMCP}}\\
\midrule
 & \multicolumn{5}{c@{\hspace{4pt}}|}{\textit{NSW}} & \multicolumn{5}{c}{\textit{QLD}}\rule[-0.9ex]{0pt}{0pt}\\
Coverage & \makebox[3.5em][r]{0.905} & \makebox[3.5em][r]{0.905} & \makebox[3.5em][r]{0.899} & \makebox[3.5em][r]{0.890} & \makebox[3.5em][r]{0.889} & \makebox[3.5em][r]{0.908} & \makebox[3.5em][r]{0.907} & \makebox[3.5em][r]{0.899} & \makebox[3.5em][r]{0.888} & \makebox[3.5em][r]{0.890}\\
Mean width & \makebox[3.5em][r]{53.0} & \makebox[3.5em][r]{59.4} & \makebox[3.5em][r]{55.9} & \makebox[3.5em][r]{52.8} & \makebox[3.5em][r]{47.2} & \makebox[3.5em][r]{51.1} & \makebox[3.5em][r]{57.2} & \makebox[3.5em][r]{$\infty$} & \makebox[3.5em][r]{50.5} & \makebox[3.5em][r]{43.1}\\
Median width & \makebox[3.5em][r]{49.6} & \makebox[3.5em][r]{57.2} & \makebox[3.5em][r]{50.1} & \makebox[3.5em][r]{52.3} & \makebox[3.5em][r]{46.5} & \makebox[3.5em][r]{43.6} & \makebox[3.5em][r]{48.5} & \makebox[3.5em][r]{40.9} & \makebox[3.5em][r]{49.1} & \makebox[3.5em][r]{38.0}\\
Winkler & \makebox[3.5em][r]{89.1} & \makebox[3.5em][r]{96.6} & \makebox[3.5em][r]{94.7} & \makebox[3.5em][r]{94.5} & \makebox[3.5em][r]{90.2} & \makebox[3.5em][r]{102.3} & \makebox[3.5em][r]{107.2} & \makebox[3.5em][r]{$\infty$} & \makebox[3.5em][r]{115.1} & \makebox[3.5em][r]{100.2}\\
MSIS & \makebox[3.5em][r]{3.88} & \makebox[3.5em][r]{4.21} & \makebox[3.5em][r]{4.12} & \makebox[3.5em][r]{4.14} & \makebox[3.5em][r]{3.93} & \makebox[3.5em][r]{6.17} & \makebox[3.5em][r]{6.38} & \makebox[3.5em][r]{$\infty$} & \makebox[3.5em][r]{6.75} & \makebox[3.5em][r]{5.99}\\
\midrule
 & \multicolumn{5}{c@{\hspace{4pt}}|}{\textit{SA}} & \multicolumn{5}{c}{\textit{TAS}}\rule[-0.9ex]{0pt}{0pt}\\
Coverage & \makebox[3.5em][r]{0.875} & \makebox[3.5em][r]{0.898} & \makebox[3.5em][r]{0.899} & \makebox[3.5em][r]{0.880} & \makebox[3.5em][r]{0.888} & \makebox[3.5em][r]{0.897} & \makebox[3.5em][r]{0.908} & \makebox[3.5em][r]{0.898} & \makebox[3.5em][r]{0.887} & \makebox[3.5em][r]{0.890}\\
Mean width & \makebox[3.5em][r]{101.1} & \makebox[3.5em][r]{109.0} & \makebox[3.5em][r]{$\infty$} & \makebox[3.5em][r]{97.0} & \makebox[3.5em][r]{87.7} & \makebox[3.5em][r]{31.8} & \makebox[3.5em][r]{34.0} & \makebox[3.5em][r]{$\infty$} & \makebox[3.5em][r]{39.7} & \makebox[3.5em][r]{29.7}\\
Median width & \makebox[3.5em][r]{87.1} & \makebox[3.5em][r]{94.2} & \makebox[3.5em][r]{102.9} & \makebox[3.5em][r]{98.0} & \makebox[3.5em][r]{85.2} & \makebox[3.5em][r]{30.6} & \makebox[3.5em][r]{32.2} & \makebox[3.5em][r]{30.6} & \makebox[3.5em][r]{39.7} & \makebox[3.5em][r]{29.0}\\
Winkler & \makebox[3.5em][r]{195.3} & \makebox[3.5em][r]{180.9} & \makebox[3.5em][r]{$\infty$} & \makebox[3.5em][r]{212.3} & \makebox[3.5em][r]{186.1} & \makebox[3.5em][r]{53.6} & \makebox[3.5em][r]{53.5} & \makebox[3.5em][r]{$\infty$} & \makebox[3.5em][r]{71.9} & \makebox[3.5em][r]{52.3}\\
MSIS & \makebox[3.5em][r]{3.14} & \makebox[3.5em][r]{2.91} & \makebox[3.5em][r]{$\infty$} & \makebox[3.5em][r]{3.45} & \makebox[3.5em][r]{3.00} & \makebox[3.5em][r]{3.81} & \makebox[3.5em][r]{3.81} & \makebox[3.5em][r]{$\infty$} & \makebox[3.5em][r]{5.11} & \makebox[3.5em][r]{3.72}\\
\midrule
 & \multicolumn{5}{c@{\hspace{4pt}}|}{\textit{VIC}} & \multicolumn{5}{c}{}\rule[-0.9ex]{0pt}{0pt}\\
Coverage & \makebox[3.5em][r]{0.898} & \makebox[3.5em][r]{0.906} & \makebox[3.5em][r]{0.899} & \makebox[3.5em][r]{0.886} & \makebox[3.5em][r]{0.889} & \makebox[3.5em][r]{} & \makebox[3.5em][r]{} & \makebox[3.5em][r]{} & \makebox[3.5em][r]{} & \makebox[3.5em][r]{}\\
Mean width & \makebox[3.5em][r]{41.0} & \makebox[3.5em][r]{45.4} & \makebox[3.5em][r]{$\infty$} & \makebox[3.5em][r]{46.6} & \makebox[3.5em][r]{37.7} & \makebox[3.5em][r]{} & \makebox[3.5em][r]{} & \makebox[3.5em][r]{} & \makebox[3.5em][r]{} & \makebox[3.5em][r]{}\\
Median width & \makebox[3.5em][r]{38.5} & \makebox[3.5em][r]{42.0} & \makebox[3.5em][r]{40.4} & \makebox[3.5em][r]{46.8} & \makebox[3.5em][r]{37.7} & \makebox[3.5em][r]{} & \makebox[3.5em][r]{} & \makebox[3.5em][r]{} & \makebox[3.5em][r]{} & \makebox[3.5em][r]{}\\
Winkler & \makebox[3.5em][r]{72.7} & \makebox[3.5em][r]{75.7} & \makebox[3.5em][r]{$\infty$} & \makebox[3.5em][r]{86.5} & \makebox[3.5em][r]{72.5} & \makebox[3.5em][r]{} & \makebox[3.5em][r]{} & \makebox[3.5em][r]{} & \makebox[3.5em][r]{} & \makebox[3.5em][r]{}\\
MSIS & \makebox[3.5em][r]{1.75} & \makebox[3.5em][r]{1.83} & \makebox[3.5em][r]{$\infty$} & \makebox[3.5em][r]{2.09} & \makebox[3.5em][r]{1.75} & \makebox[3.5em][r]{} & \makebox[3.5em][r]{} & \makebox[3.5em][r]{} & \makebox[3.5em][r]{} & \makebox[3.5em][r]{}\\
\bottomrule
\end{tabular}}

}

\end{table}%

The comparison across regions confirms that the relative performance of
the five methods depends on the forecast horizon and on the evaluation
measure. AcMCP has the lowest mean interval width in each region while
maintaining empirical coverage close to the nominal level. Its advantage
is less consistent for the Winkler and MSIS scores, however, as SCP and
WCP perform better in some regions. Thus, the narrower intervals
produced by AcMCP do not always translate into the best overall interval
score.

The results for ACP require separate attention because this method
produces unbounded intervals in Queensland, South Australia, Tasmania,
and Victoria. Under the ACP update, repeated tail violations can drive a
tail-specific miscoverage level toward its lower boundary, at which
point the corrected quantile may become infinite. Such intervals remain
in the empirical coverage calculation and lead to infinite mean width,
Winkler, and MSIS values. The package records these values as
\texttt{Inf} rather than excluding them, ensuring that their effect
remains visible in mean-based measures, while the finite median widths
show that the typical ACP interval remains comparable in size to those
produced by the other methods.

\section{Conclusion}\label{sec-conclusion}

This paper has presented \textbf{conformalForecast}, an R package for
constructing, updating, and evaluating conformal prediction intervals
for multistep time series forecasts. Its design follows from four
requirements of sequential forecasting. Point forecasts may come from
any source, so calibration is kept separate from forecasting. Forecast
errors depend on both origin and horizon, so they are stored in a common
origin--horizon representation on which the five calibration schemes
operate through a single interface. Observations keep arriving, so
\texttt{update()} extends a fitted object from its stored score
histories and recursive states instead of rebuilding it. Performance
varies over time and across horizons, so \texttt{coverage()},
\texttt{width()}, and \texttt{accuracy()} report it in aggregate, on
rolling windows, and by horizon.

In the electricity price application, pooled coverage is close to 90\%
for all methods, while the rolling and horizon-specific summaries reveal
variation that the aggregate conceals. AcMCP gives the narrowest
intervals on average in all five regions, but not always the best
Winkler score or MSIS, and ACP issues unbounded intervals in four
regions. No method is best on every criterion, and these rankings are
specific to the data and forecasting setup. Updating the fitted objects
reproduces the intervals obtained by rebuilding them at a small fraction
of the computational cost, because only the newly arrived origins are
processed.

The common interface standardizes how the methods are applied and
evaluated. It does not make them interchangeable, since each retains its
own assumptions and coverage statement. What it changes is the setting
in which they can be used. Sequential conformal methods are typically
described as algorithms acting on a single stream of forecasts and
outcomes. In practice, forecasts are issued at successive origins for
several horizons, by models that are chosen and maintained separately
from any calibration step, and they are revised as new data arrive. By
building its workflow around forecast origins, horizons, and arriving
data, \textbf{conformalForecast} brings these methods into that setting
and makes them usable where forecasts are produced and revised.

Several limitations also indicate directions for future development.
\textbf{conformalForecast} currently uses conventional R vectors,
matrices, and S3 objects and does not yet provide native
tidyverse/tidyverts workflows. This framework-neutral core is
intentional: it keeps calibration independent of a particular data
grammar and allows forecasts from different models and external systems
to share the same origin--horizon representation. Adapters for
\texttt{tsibble} and \texttt{fable} objects could nevertheless improve
interoperability without altering this core. The package also implements
a focused subset of the expanding conformal prediction literature.
Future releases could add methods compatible with sequential multistep
calibration and incremental updating, together with principled defaults,
diagnostics, and data-adaptive recommendations for calibration windows,
decay weights, learning rates, and PID or scorecaster settings. Work on
online aggregation and adaptive step-size selection provides useful
starting points \citep{zaffran2022, gibbs2024, Liu2026-expert}, although
any tuning procedure should preserve the information requirements and
coverage statement of the underlying method.

\section*{Declaration of generative AI
use}\label{declaration-of-generative-ai-use}
\addcontentsline{toc}{section}{Declaration of generative AI use}

During the preparation of this work the authors used ChatGPT and Claude
to improve English language and assist in coding to create graphical
illustrations. After using these tools, the authors reviewed and edited
the content as needed and take full responsibility for the content of
the publication.

\section*{References}\label{references}
\addcontentsline{toc}{section}{References}

\renewcommand{\bibsection}{}
\bibliography{references}

\end{document}